# Modelling Human Active Search in Optimizing Black-box Functions

Antonio Candelieri[1][0000-0003-1431-576X], Riccardo Perego[1][0000-0003-0117-2237], Ilaria Giordani[1][0000-0002-6065-0473], Andrea Ponti and Francesco Archetti[1][0000-0003-1131-3830]

[1] University of Milano-Bicocca, Milan 20126, Italy
antonio.candelieri@unimib.it

**Abstract.** Modelling human function learning has been the subject of intense research in cognitive sciences. The topic is relevant in black-box optimization where information about the objective and/or constraints is not available and must be learned through function evaluations. In this paper we focus on the relation between the behaviour of humans searching for the maximum and the probabilistic model used in Bayesian Optimization. As surrogate models of the unknown function both Gaussian Processes and Random Forest have been considered: the Bayesian learning paradigm is central in the development of active learning approaches balancing exploration/exploitation in uncertain conditions towards effective generalization in large decision spaces. In this paper we analyse experimentally how Bayesian Optimization compares to humans searching for the maximum of an unknown 2D function. A set of controlled experiments with 60 subjects, using both surrogate models, confirm that Bayesian Optimization provides a general model to represent individual patterns of active learning in humans.

## 1    Introduction

We consider as reference problem the black-box optimization: the objective function and/or constraints are analytically unknown and evaluating them might be very expensive and noisy. In black-box situations as we cannot assume any prior knowledge about the objective function $f(x)$, any functional form is a priori admissible and the value of the function at a point says nothing about the value at other points: the only way to develop a problem specific algorithm is to assume a model of $f(x)$ and to learn through a sample of function values.

Such an algorithm must be sample efficient, because the cost of function evaluations is the dominating cost. This problem has been addressed in several fields under different names, including active learning (Kruschke et al., 2008; Griffiths et al., 2008; Wilson et al., 2015), Bayesian Optimization (BO) (Zhigljavsky and Zilinskas, 2007), (Candelieri et al., 2018), (Archetti et al. 2019), hyperparameter optimization (Eggensperger et al., 2019) and others.

In the BO framework a surrogate model of the objective function is built to sum up our a priori beliefs about the objective function and the informative value of new observations. Two probabilistic frameworks are usually considered: the Gaussian Processes (GPs) and Random Forests (RF) which offer alternative ways to update the beliefs as new data arrives and to provide an estimate of the expected value of the objective function and the uncertainty in this estimate.

A distinction is usually drawn among them accordingly to the type of design variables: continuous ones are better dealt with GP while integer/categorical and conditional ones with RF.

In both cases the next sampled point is chosen on the basis of its informative value through the maximization of an acquisition function (also called infill): this choice brings up the so called "exploration vs exploitation dilemma", where exploration means devoting resources to know more about possible solutions while exploitation devotes resources to improve on solutions already identified in previous phases. The search for the new point must strike an effective balance between the needs of exploration and exploitation.

Psychologists have extensively studied how humans balance exploration and exploitation (Krusche et al., 2008), (Mehlhorn et al., 2015), with a recent attention on the links between modern machine learning algorithms and psychological processes. (Gershman, 2018; Schulz et al., 2016; Gopnik et al., 2017). Psychological research has mostly focused on how people learn functions according to a protocol in which an input is presented to participants and they are asked to predict the corresponding output value. Then they observe the true output value (usually noisy) in order to update their own "predictive model" that is to adjust their internal representation of the underlying function. Psychologists have largely focused on GP: the issue of GP regression, kernel composition for different degrees of smoothness and safe optimization in their relation to cognition is studied in a recent survey by (Shultz et al., 2018). Directed exploration is realized by adding the so-called *uncertainty bonuses* to estimated values obtaining the *upper confidence bound* (UCB) algorithm (Srinivas et al., 2010). In (Wu et al., 2018) the human search strategy is analysed for rewards under limited search horizons, concluding that GP offers the best model for generalization and UCB the best solution of the exploration/exploitation dilemma.

A significant application of RF is given in (Plonsky et al., 2019) as a hybrid model of machine learning and decision mechanisms. A key driver in the above research activities is that Human learners are increasingly fast at adapting to unfamiliar environments. Psychologists are investigating the intriguing gap between the capabilities of human and machine learning.

Most previous research findings in human learning refer to function learning because is related to a probabilistic perspective on predictability and provides a proxy to generalization capability. Contrary to function learning, optimization is not yet widely considered in the literature; in (Borji & Itti, 2013) a simple 1-D optimization problem has been considered.

The approach presented in this paper has been sketched in (Archetti et al., 2019). The present paper has been significantly augmented and rewritten. A set of new computational results are related to the use, along with the GP, of the RF as surrogate model. The set of references has been also enlarged and the whole perspective has been widened to reflect that: learning and optimization of black-box functions are two faces of the same coin. GP and RF are shown to offer a reasonable unifying framework of human function learning, active sampling and optimization.

The structure of the paper is as follows: section 2 outlines the methodological background of BO including the basis of the 2 main surrogate models GP and RF and the management of the exploration/exploitation dilemma. Section 3 is devoted to the experimental set-up and section 4 reports the experimental results about the behavioural patterns of humans in optimizing black box functions. Section 5 outlines the conclusions of this study and the perspectives of future works.

## 2 Methodological background

This section provides the underlying methodological framework of the study. The global optimization problem we consider is defined as:

$$\max_{x \in X \subset \mathbb{R}^d} f(x)$$

where the search space $X$ is generally box-bounded and $f(x)$ is black-box meaning that no gradient information is available and that we have only access to noisy observation of $f$ which are computationally expensive.

### 2.1 Gaussian Processes

GPs are a powerful non-parametric model for implementing both regression and classification. One way to interpret a GP is as a distribution over functions, with inference taking place directly in the space of functions (Williams and Rasmussen, 2006). A GP, therefore, is a collection of correlated random variables, any finite number of which have a joint Gaussian distribution. A GP is completely specified by its mean function $\mu(x)$ and covariance function $cov\big(f(x), f(x')\big) = k(x, x')$:

$$\mu(x) = \mathbb{E}[f(x)]$$

$$cov\big(f(x), f(x')\big) = k(x, x') = \mathbb{E}[(f(x) - \mu(x))(f(x') - \mu(x'))]$$

and will be denoted by: $f(x) \sim GP\big(\mu(x), k(x, x')\big)$. This means that the behaviour of the model can be controlled entirely through the mean and covariance.

Usually, for notational simplicity we will take the prior of the mean function to be zero, although this is not necessary. The covariance function assumes a critical role int the GP modelling, as it specifies the distribution over functions, depending on a sample $X_{1:n} = \{x_1, \dots, x_n\}$ with $f(X_{1:n}) \sim \mathcal{N}\big(\mathbf{0}, \mathrm{K}(X_{1:n}, X_{1:n})\big)$ and $\mathrm{K}(X_{1:n}, X_{1:n})$ is a n×n matrix whose entry $\mathrm{K}_{ij} = \mathrm{k}(x_i, x_j)$ thus the covariance specifies how to points a correlated and it controls the shape of the objective function.

We usually have access only to noisy function values, denoted by $y = f(x) + \varepsilon$. Assuming additive independent identically distributed Gaussian noise $\varepsilon$ with variance $\lambda^2$ and let $y = (y_1, \ldots, y_n)$ whose covariance is $\mathrm{K}(X_{1:n}, X_{1:n}) + \lambda^2 I$.

Let $D_{1:n} = \{(x_i, y_i)\}_{i=1,\ldots,n}$, where $y_i = f(x_i) + \varepsilon$, and $\varepsilon \sim \mathcal{N}(0, \lambda^2)$ in the case of a noisy objective function.

Therefore, the predictive equations for GP regression, that are $\mu(x)$ and $k(x, x')$, can be easily updated, by conditioning the joint Gaussian prior distribution on the observations:

$$\mu(x) = \mathbb{E}[f(x)|D_{1:n}, x] = \mathrm{k}(x, X_{1:n})[\mathrm{K}(X_{1:n}, X_{1:n}) + \lambda^2 I]^{-1} y$$

$$\sigma^2(x) = k(x, x) - \mathrm{k}(x, X_{1:n})[\mathrm{K}(X_{1:n}, X_{1:n}) + \lambda^2 I]^{-1} \mathrm{k}(X_{1:n}, x)$$

where $\mathrm{K}(X, X_{1:n}) = [k(x, x_1), \ldots, k(x, x_n)]$.

The covariance function is the crucial ingredient in a GP predictor, as it encodes assumptions about the function to approximate: function evaluations that are near to a given point should be informative about the prediction at that point. Under the GP view it is the covariance function that defines nearness or similarity. Once the prior mean and the kernel are chosen, they are updated with the observation of $f$ to find a-posterior distribution $f(x \mid D_{1:n})$ and this allows us to find the expected value of the function at any point and to calculate its uncertainty through its predicted variance.

Examples of covariance (aka kernel) functions:

*Squared Exponential (SE) kernel:*

$$k_{SE}(x, x') = e^{-\frac{\|x - x'\|^2}{2\ell^2}}$$

with $\ell$ known as *characteristic length-scale*.

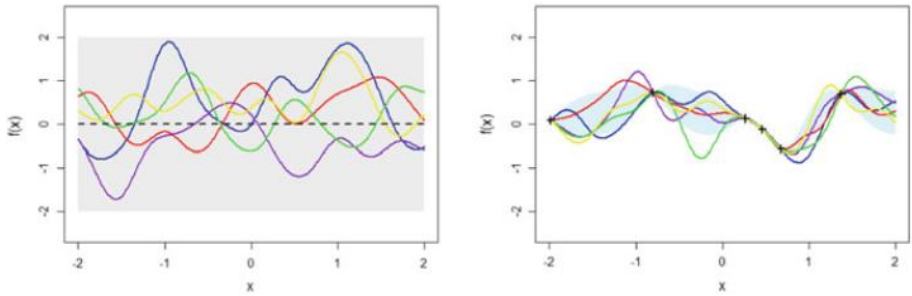

*Exponential kernel:*

$$k_{Exp}(x, x') = e^{-\frac{|x - x'|}{\ell}}$$

*Power Exponential kernel:*

$$k_{PowExp}(x, x') = e^{-\left(\frac{|x-x'|}{\ell}\right)^p}$$

*Matérn kernels:*

$$k_{Mat}(x, x') = \frac{2^{1-\nu}}{\Gamma(\nu)} \left(\frac{|x-x'|\sqrt{2\nu}}{\ell}\right)^\nu K_\nu \left(\frac{|x-x'|\sqrt{2\nu}}{\ell}\right)$$

Where $\ell$ is the length scale (updated by maximum likelihood destination) and $\nu > 0$ is a parameter governing the regularity of the gp samples which are $\nu - 1$ differential and where $\Gamma(\nu)$ is the gamma function and $K_\nu$ is the modified Bessel function of the second kind. The most widely adopted versions, specifically in the Machine Learning community and considered in this paper, are $\nu = 3/2$ and $\nu = 5/2$. The Matern kernel encodes the expected smoothness of the target function explicitly.

## 2.2 Random Forest

Random Forest (RF) is an ensemble learning method, based on decision trees, for both classification and regression problems (Ho, 1995). According to the originally proposed implementation, RF aims at generating a multitude of decision trees, at training time, and providing as output the mode of the classes (classification) or the mean/median prediction (regression) of the individual trees.

Although originally designed and presented as a machine learning algorithm, RF is also an effective and efficient alternative to GP for implementing BO. RF consists of an ensemble of different regressors (i.e., decision trees), it is possible to compute—as for GP – both μ(x) and σ(x), simply as mean and variance of the samples of the individual outputs provided by the regressors. Due to the different nature of RF and GP, the associated probabilistic surrogate models will also result significantly different. While GP is well-suited to model smooth functions in search space spanned by continuous variables, RF can also deal with discrete and conditional variables.

A basic description of the RF learning algorithm is provided in 3.2. Albeit GPs offer better mathematical characterization, RFs often result more computationally efficient than GP, (even with continuous variables), also because RFs do not require to invert any kernel matrix.

## 2.3 The acquisition functions

The acquisition function is the mechanism to implement the trade-off between exploration and exploitation in BO. More precisely, any acquisition function aims to guide the search of the optimum towards points with potentially high values of objective function either because the prediction of $f(x)$, based on the probabilistic surrogate model, is high or the uncertainty, also based on the same model, is high (or both). Indeed, exploitation means to consider the area providing more chance to improve over current solution, while exploration means to move towards less explored regions of the search space where predictions based on the surrogate model are more uncertain, with higher variance. There are many acquisition functions, we quote only those used in this study. *Probability of Improvement* (PI) (Kushner 1964) and *Expected Improvement* (EI) (Mockus 1975) measure, respectively, the probability and the expectation of the improvement over the best observed value of $f(x)$ given the predictive distribution of the probabilistic surrogate model. More recently, *Upper/Lower Confidence Bound*, (Srinivas et al., 2010) is widely used. It is an acquisition function that manages exploration-exploitation by being optimistic in the face of uncertainty where Upper and Lower are used, respectively, for maximization and minimization problems:

$$\text{LCB}(x) = \mu(x) - \xi\sigma(x)$$

$$\text{UCB}(x) = \mu(x) + \xi\sigma(x)$$

where $\xi \geq 0$ is the parameter to manage the trade-off between exploration and exploitation ($\xi = 0$ is for pure exploitation; on the contrary, higher values of $\xi$ emphasizes exploration by inflating the model uncertainty). In (Srinivas et al., 2010), a policy is provided for updating the value of $\xi$ along function evaluations, with also a proof of convergence of such a policy.

In the case of a minimization problem the next point is chosen as

$$x_{n+1} = \underset{x \in X}{\operatorname{argmin}} \, LCB(x)$$

while, in the case of a maximization problem the next point is selected as

$$x_{n+1} = \underset{x \in X}{\operatorname{argmax}} \, UCB(x)$$

### 2.4 Bayesian optimization

The following algorithm summarizes a general Bayesian Optimization process where the acquisition function, whichever it is, is denoted by $\alpha(x, D_{1:n})$. This function is generally maximized, but in the case of $\alpha = LCB$.

With respect to the probabilistic surrogate model, the summarized algorithm does not specify the probabilistic surrogate model, as well as the kernel in the case of a GP. This is basically done in order to maintain the algorithm as general as possible.

In this study we have used GP (considering the kernel presented in the previous section) and RF as surrogate probabilistic models. The three different acquisition functions previously described have been used for both the two surrogates.

---

**General Bayesian Optimization Algorithm**

---

Generate an initial set of $m$ points $X_{1:m}$ randomly sampled (e.g., via Latin Hypercube Sampling)

Evaluate the function in the initial set of points and obtain $D_{1:m}$

Define a further budget $N$

1 **for** $n = m, ..., m + N$ **do**

2 update the surrogate model obtaining the new estimates of $\mu(x)$ and $\sigma(x)$

3 select a new $x_{n+1}$ by optimizing an acquisition function $\alpha$, such that

$$x_{n+1} = \underset{x}{\operatorname{argmax}} \, \alpha(x | D_{1:n})$$

4 evaluate the objective function to obtain $y_{n+1} = f(x_{n+1})$

5 update the dataset of observations $D_{1:n+1} = D_{1:n} \cup \{(x_{n+1}, y_{n+1})\}$

6 **endfor**

7 Output: the best $y$ value observed over the entire optimization process

---

## 3      Experimental setup

### 3.1      User interface

Stimuli are 15 different 2D functions among which at the start of each game the test function is randomly chosen (Adorio et al., 2005). Each subject was informed about the goal of the experiment and the available number of clicks for the play, before to start. Subjects started by clicking any point in the screen getting the corresponding value; previously clicked points remained on the screen until the end of the trial. In particular, points are colored and resized according to the associated score, providing a visual feedback about the distribution of the scores collected so far.

We have developed three different game modalities:

1. Find the point with maximum value without knowing its value. In this case the humans have to optimize the simple regret (without knowing $f(x^*)$):

$$\min_{n=1,...,N} f(x^*) - y_n^+$$

   where $y_n^+ = \max_{i=1,...,n} y_i$

2. Find the point with maximum value knowing its value. As in the previous modality, the humans have to optimize the simple regret, but without knowing $f(x^*)$

3. Maximize the total score, that is the cumulative value of the selected points, without knowing the value of the maximum. In this case the humans have to optimize the cumulative regret (without knowledge):

$$\min \sum_{i=1}^{N} (f(x^*) - y_i)$$

Data of each game are stored in a database with the following structure (Fig. 1). In games table, each row represents a single point of a game, specifying the user, the function and the game mode. The games are identified by the timestamp relative to the end of the game.

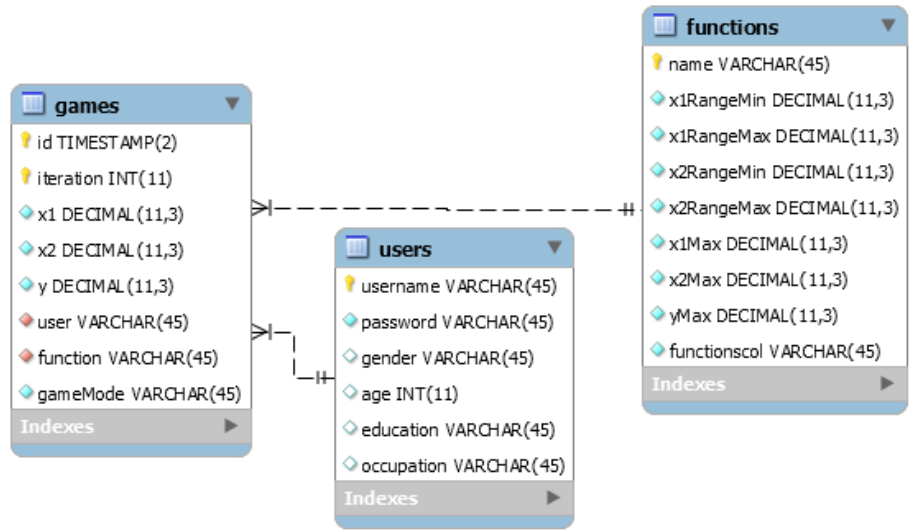

**Fig. 1.** ER diagram of database

The following picture (Fig. 2) shows a frame of the game.

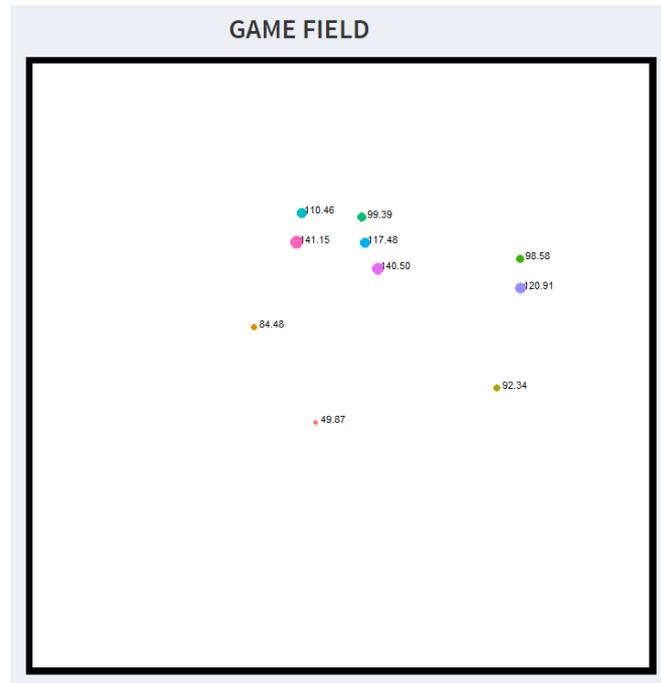

**Fig. 2.** An example of a game play

### 3.2 Procedure

For each player, at each iteration, a GP and RF models are fitted on the observed points and the three acquisition functions, previously mentioned in section 2, are used to select the next point to query. All these points are compared, via Euclidean distance, to the corresponding choice made by the player.

Moreover, for the GP, five kernels have been used to consider different possibility of smoothness approximation of the objective function.

Human player choices and Bayesian Optimization are considered compliant, pointwise, if the distance between the point chosen by the human player and the algorithmic player is less than a given "threshold". Finally, the strategy of the human player is assimilated to the acquisition function most frequently compliant, pointwise, along a play. The procedures for GP-based and RF-based BO are summarized in the following pseudo-codes:

---

**Algorithm for compliance analysis with respect to GP-based BO**

Let denote by:
  - $p$ a participant
  - $k$ a kernel
  - $\alpha$ an acquisition function
  - $n$ a generic iteration

- $s^{p,k,n}$ a search strategy (i.e., an acquisition fuction, under a GP with a given kernl, at a specific iteration)

1 **foreach** $p$, $k$ and $n$
2 fit a GP $GP^{p,k,n}$
3 **foreach** $\alpha$
4 select $x_{n+1}^{GP_{p,k,n}}$
5 compute $d^{p,k,n,\alpha} = \left\| x_{n+1}^{GP^{p,k,n}} - x_{n+1}^{p} \right\|$, where

$\quad$ $x_{n+1}^{GP^{p,k,n}} = \underset{x}{\mathrm{argmax}}\, \alpha(x)|GP^{p,k,n}$ is the next point according to acquisition
$\quad$ function $\alpha$ under $GP^{p,k,n}$
$\quad$ and $x_{n+1}^{p}$ is the next point chosen by participant $p$ at iteration $n+1$
6 **endforeach**
7 $\bar{d}^{p,k,n} = \underset{\alpha}{\min}\{d^{p,k,n,\alpha}\}$
8 **if** $\bar{d}^{p,k,n} \leq threshold$ **then**
9 $s^{p,k,n} = \underset{\alpha}{\mathrm{argmin}}\{d^{p,k,n,\alpha}\}$
10 **else**
11 $s^{p,k,n} = \emptyset$

Finally, for a given kernel $\bar{k}$, the search strategy of the participant $\bar{p}$ is compliant to the most frequent acquisition function in the series $s^{p,k} = \{s^{p,k,n}\}_{n=1:N}$.

**Algorithm for compliance analysis with respect to RF-based BO**

Let denote by:
- $p$ a participant
- $\alpha$ an acquisition function
- $n$ a generic iteration
- $s^{p,n}$ a search strategy (i.e., an acquisition function)

1 **foreach** $p$ and $n$
2 fit a Random Forest $RF^{p,n}$
3 **foreach** $\alpha$
4 select $x_{n+1}^{RF_{p,n}}$
5 compute $d^{p,n,\alpha} = \left\| x_{n+1}^{RF^{p,n}} - x_{n+1}^{p} \right\|$, where

$\quad$ $x_{n+1}^{RF^{p,n}} = \underset{x}{\mathrm{argmax}}\, \alpha(x)|RF^{p,n}$ is the next point according to acquisition func-
$\quad$ tion $\alpha$ under $RF^{p,n}$
$\quad$ and $x_{n+1}^{p}$ is the next point chosen by participant $p$ at iteration $n+1$
6 **endforeach**
7 $\bar{d}^{p,n} = \underset{\alpha}{\min}\{d^{p,n,\alpha}\}$
8 **if** $\bar{d}^{p,n} \leq threshold$ **then**
9 $s^{p,n} = \underset{\alpha}{\mathrm{argmin}}\{d^{p,n,\alpha}\}$
10 **else**

Finally, the search strategy of the participant $\bar{p}$ is compliant to the most frequent acquisition function in the series $s^p = \{s^{p,n}\}_{n=1:N}$.

### 3.3 Software resources and analysis

Software resources consists of a pipeline of two components developed in R. The first component is the procedure to compute the distance between the humans' and the BO's choices during each game, the second aggregates all the calculated distances and generates the related statistics.

The first component uses the R package named "mlrMBO" (https://cran.r-project.org/web/packages/mlrMBO/vignettes/mlrMBO.html). This library offers both GP and RF as surrogate model along with the acquisition functions adopted in this study. L-BFGS algorithm is used to optimize the acquisition function based on GP and with continuous search space; otherwise, "focussearch" algorithm (Bischl et al. 2017) is adopted: it can handle with numeric, discrete and mixed search spaces, also involving conditional variables. Focus-search starts with a large set of random points where the acquisition function is evaluated. Then, it shrinks the search space around the current best point and perform a new random sampling of points within the "focused space". The shrinkage operation is iteratively performed until a maximum number of iterations and the entire procedure can be restarted multiple times to mitigate the risk to converge to a local optimum. Finally, the best point over all restarts and iterations is returned as the solution.

## 4 Experimental results

### 4.1 Experiment 1 – Gaussian Process

According to the mentioned procedure, the following figures summarize the main results of the study.

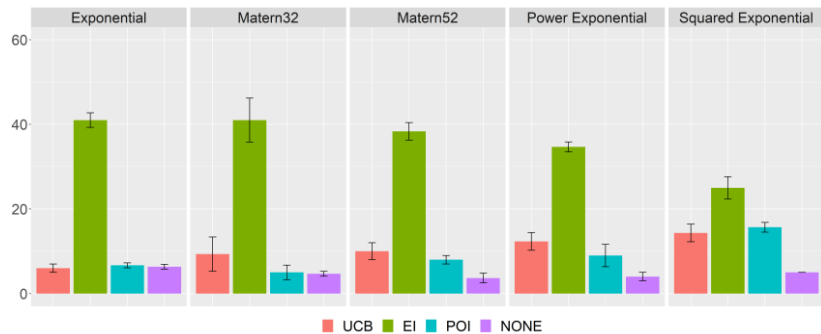

**Fig. 3.** Number of human players whose strategy is compliant with respect to kernel type and acquisition functions, with "threshold" set to 0.10, and $\beta = 1$ in UCB. Last bar represents the number of non-compliant.

From Fig. 3, EI is preferred, indicating that exploitative behaviour is dominant among humans. This outcome is relatively independent on the kernel.

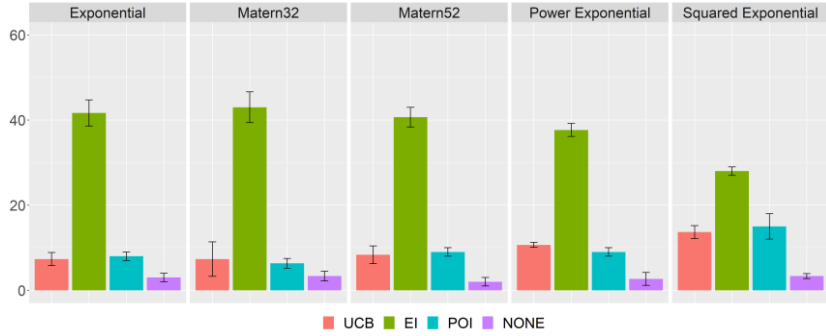

**Fig. 4.** Number of human players whose strategy is compliant with respect to kernel type and acquisition functions, with "threshold" set to 0.15, and $\beta = 1$ in UCB. Last bar represents the number of non-compliant.

From Fig. 4, results are coherent with previous Fig. 3. Moreover, the number of non-compliant is reduced with the less restrictive threshold.

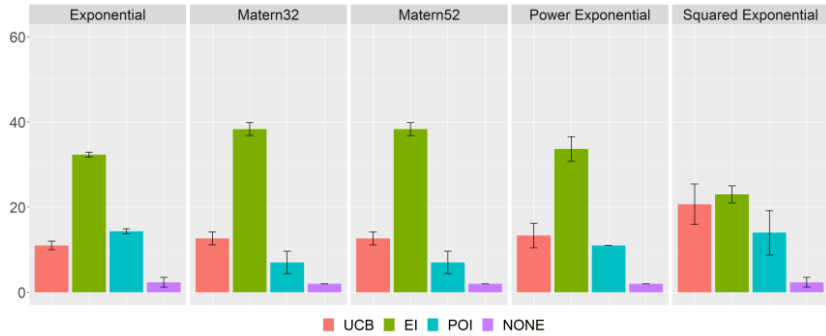

**Fig. 5.** Number of human players whose strategy is compliant with respect to kernel type and acquisition functions, with "threshold" set to 0.15, and $\beta = 0.5$ in UCB. Last bar represents the number of non-compliant.

From Fig. 5, one can see that with the reduction in $\beta$, UCB gains a larger share of participants than previously.

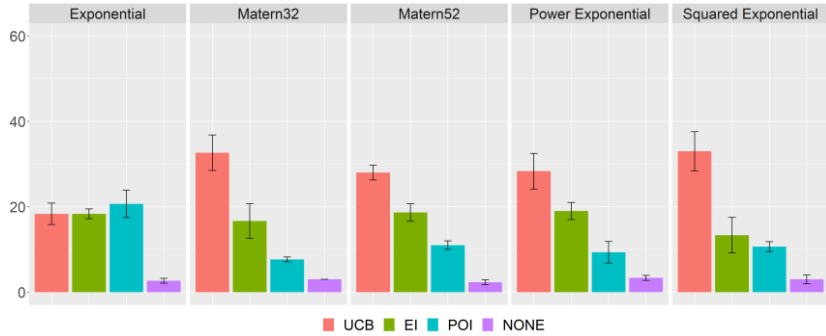

**Fig. 6.** Number of human players whose strategy is compliant with respect to kernel type and acquisition functions, with "threshold" set to 0.15, and $\beta = 0$ in UCB. Last bar represents the number of non-compliant.

From Fig. 6, the last confirmation that "greed is good": $\beta = 0$ means no-exploration in UCB whose fully exploitative version gets the largest share of participants.

### 4.2    Experiment 2 – Random Forest

From Fig. 7, with and explorative UCB ($\beta = 1$), Probability of Improvement, a notoriously exploitative acquisition function, gets the larger share. The situation changes reducing the exploration component in UCB, which with $\beta = 0.5$ and $\beta = 0.0$ dominates choices among the compliant.

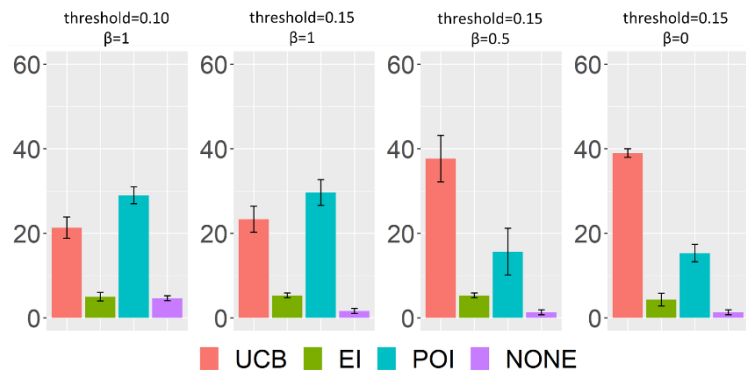

**Fig. 7** Number of human players whose strategy is compliant with respect to different acquisition functions, with RF, in different settings of "threshold" and $\beta$ values of UCB. From left to right: threshold = 0.10 and $\beta = 1$; threshold = 0.15 and $\beta = 1$, $\beta = 0.5$ and $\beta = 0$.

## 5      Conclusions

BO is a principled approach for adding a mathematical structure to the search and optimization process which resembles human optimization strategies. This can be explained by the fact that probabilistic surrogate models explain function approximation in humans. In order to search efficiently for the optimum, one needs to learn the function landscape by updating its approximation through observations. Indeed, learning and optimization of black-box functions are 2 faces of the same coin.

GP and RF have been argued to offer a reasonable unifying framework of human function learning, efficient active sampling and search. While most research findings have been focused on human errors in function learning we focus on optimization of black-box functions, linked to human active search behavior.

The number of BO compliant participants is very high (less than 10% of participants resulted non-compliant to all models and acquisition functions). A general conclusion is that the exploitative oriented acquisition functions are get consistently the larger share. Also interesting is the analysis of which space model, that is kernel, and which exploitation-exploration balance, that is the acquisition function, are implied by human search. Based on the limited set of results presented in this paper, kernel is not a major factor in determining compliance. The acquisition function, and its parametrization for UCB, are the main determinants of the choice. The value of $\beta$ significantly affects the relative importance of UCB over the search processes performed by the compliant participants. Finally, we can conclude that "greed is good", that is the human behavior is quite exploitative (i.e., threshold=0.15 and $\beta$=0 for UCB).

## 6      Compliance with ethical standards

Ethical approval: All procedures performed in studies involving human participants were in accordance with the ethical standards of the institutional and/or national research committee and with the 1964 Helsinki declaration and its later amendments or comparable ethical standards.

Informed consent: Informed consent was obtained from all individual participants included in the study.